\documentclass[12pt,showpacs,prd]{revtex4}
\usepackage{epsfig,amssymb,amsfonts}

\makeatletter
\renewcommand{\@makefntext}[1]{\parindent=1em\noindent\hbox to 1.8em{\hss$^{\@thefnmark}$}#1}
\renewcommand{\@footnotemark}{\hbox{\mathsurround=0pt$^{\@thefnmark}$}}
\newcommand{\ftnote}[2]{\footnotemark[#1]\footnotetext[#1]{#2}}
\makeatother

\begin{document}
\title{Chiral symmetry breaking and the Lorentz nature of confinement}
\author{A. V. Nefediev}
\affiliation{Institute of Theoretical and Experimental Physics, 117218,\\
B.Cheremushkinskaya 25, Moscow, Russia}
\author{Yu. A. Simonov}
\affiliation{Institute of Theoretical and Experimental Physics, 117218,\\
B.Cheremushkinskaya 25, Moscow, Russia}
\newcommand{\be}{\begin{equation}}
\newcommand{\bea}{\begin{eqnarray}}
\newcommand{\ee}{\end{equation}}
\newcommand{\eea}{\end{eqnarray}}
\newcommand{\ds}{\displaystyle}
\newcommand{\low}[1]{\raisebox{-1mm}{$#1$}}
\newcommand{\loww}[1]{\raisebox{-1.5mm}{$#1$}}
\newcommand{\lmn}{\mathop{\sim}\limits_{n\gg 1}}
\newcommand{\vpint}{\int\makebox[0mm][r]{\bf --\hspace*{0.133cm}}}
\newcommand{\too}{\mathop{\approx}\limits_{r\gg T_g}}
\newcommand{\vp}{\varphi}
\newcommand{\vx}{\vec{x}}
\newcommand{\vy}{\vec{y}}
\newcommand{\vz}{\vec{z}}
\newcommand{\vk}{\vec{k}}
\newcommand{\vq}{\vec{q}}
\newcommand{\vpp}{\vec{p}}
\newcommand{\vn}{\vec{n}}
\newcommand{\vg}{\vec{\gamma}}
\newcommand{\ld}{\lambda}
\newcommand{\cor}{D(\tau,\ld)}

\begin{abstract}
We address the question of the Lorentz nature of the effective interquark interaction in QCD which leads to the formation of the QCD string
between colour charges.
In particular, we start from a manifestly vectorial fundamental interaction mediated by gluons and demonstrate that,
as soon as chiral symmetry is broken spontaneously, the effective interquark interaction acquires a selfconsistently generated scalar part
which is eventually responsible for the formation of the QCD string.
We demonstrate this explicitly for a heavy--light quarkonium, using the approach of the Schwinger--Dyson-type equation
and the quantum--mechanical Hamiltonian method of the QCD string with quarks at the ends.
\end{abstract}
\maketitle

\section{Introduction}

In this paper we discuss one of the long--standing problems in QCD --- namely, the problem of the Lorentz nature of the long--range
confining interquark
interaction. QCD is believed to be a stringlike theory at large distances, that is the long--range
interquark interaction is expected to be generated by an extended object --- the QCD
string. Such a phenomenological picture appears rather successful in various studies of hadronic properties.
An important step was made in the framework of the Vacuum Correlators Method (VCM) \cite{VCM} in which the Lagrangian of
the QCD string with quarks at the ends can be derived naturally starting from the fundamental QCD Lagrangian \cite{DKS}.
In addition, the nonperturbative spin--dependent forces in heavy and light quarkonia were found, following the formalism established in
Ref.~\cite{SO}. Thus, for the spin--orbit interaction, a celebrated representation,
\be
V_{SO}(r)=\left(\frac{\vec{\sigma}_q\vec{l}_q}{4m_q^2}-\frac{\vec{\sigma}_{\bar{q}}\vec{l}_{\bar{q}}}{4m_{\bar{q}}^2}\right)
\left(\frac1r\frac{\partial\varepsilon}{\partial r}+\frac2r\frac{\partial V_1}{\partial r}\right)
+\frac{1}{2m_qm_{\bar{q}}}\left(\vec{\sigma}_{\bar q}\vec{l}_q-\vec{\sigma}_{q}\vec{l}_{\bar{q}}\right)\frac1r
\frac{\partial V_2}{\partial r},
\ee
was introduced in Ref.~\cite{SO}, where $\varepsilon (r)$ is the static confining potential, and a general relation (Gromes relation \cite{Gr}) is valid,
\be
\varepsilon'+V_1'-V_2'=0.
\ee
For a purely scalar interaction, one obtains at large $r$'s:
\be
V_1'=-\varepsilon',\quad V'_2=0,
\label{2}
\ee
and this was demonstrated explicitly for the Gaussian approximation for the field correlators in Ref.~\cite{int}.
For the case of a vector confinement, for example, for the Coulomb potential, one would find:
\be
V_1'=0,\quad V_2'>0,
\ee
and the coefficient at the spin--orbit term would have the opposite sign.
Phenomenology of the heavy quarkonia spectrum favours the first possibility, Eq.~(\ref{2}),
so that one has an evidence that Nature prefers scalar interquark interaction, at least for heavy quarks. On the lattice, numerous data 
also support the first possibility (see Ref.~\cite{km0} for recent results and the vast bibliography). 
In the meantime, any quantum--mechanical approach meets severe problems with the description of another celebrated phenomenon --- spontaneous
breaking of chiral symmetry, which is known to take place in QCD. A full quantum field theory based treatment has to be exploited for this
purpose. An example of such a treatment, also based on the VCM, is given by the Schwinger--Dyson-type approach to heavy--light quarkonia 
suggested in Ref.~\cite{hlya}, and the Lorentz nature of confinement
for heavy quarks was studied in this formalism in Refs.~\cite{hl,hlya,hlus}. On the other hand, for light
quarks, vectorlike confining interaction would have resulted in the well--known Klein paradox and, thence, in problems with building the
spectrum of hadrons. No evidence for such problems exists so far. In this paper we prove that, indeed, even for light quarks, 
if chiral symmetry is broken spontaneously, the effective interquark interaction acquires a selfconsistently generated scalar part. 
This result is quite general and holds regardless of the explicit form of
the interquark kernel, suffices it is confining and thus leads to spontaneous breaking of chiral symmetry.
A link is established between the VCM and potential quark models \cite{pqm,BR0} which we refer as to Generalised Nambu--Jona-Lasinio (GNJL) 
models, the latter being widely used for studies of low--energy phenomena in QCD.
This is an important outcome of our work since the vast results obtained in the literature in the framework of
such quark models are valid for our situation as well --- what we do is approaching the same problem
from another side. Thus we demonstrate that, starting from the fundamental
QCD Lagrangian and using the Gaussian approximation for the background field correlators, one can derive a Schwinger--Dyson-type equation for
the heavy--light quarkonium which, at large interquark distances, reduces to a Diraclike equation with an effective interquark interaction
which contains a dynamically generated scalar part, as a consequence of chiral symmetry breaking. 
At the same time, a Schr{\" o}dingerlike equation with the Hamiltonian of the QCD string with quarks at the ends (in the form of the
Salpeter equation) arises naturally 
from the same Schwinger--Dyson-type equation, if the scalar interaction dominates \cite{ns0}. We conclude, therefore, that this is the dynamical 
scalar interaction responsible for the QCD string formation.

\section{Schwinger--Dyson-type equation for a heavy--light quarkonium}

In this section we consider a heavy--light quarkonium consisting of a static antiquark and a quark, whose mass is
unconstrained and can take any value (we shall be mostly interested in the case of the massless quark).
Our starting point the heavy--light Greens function $S_{q\bar{Q}}$ written in Euclidean space as \cite{hlya}
\be
S_{q\bar Q}(x,y)=\frac{1}{N_C}\int D{\psi}D{\psi^\dagger}DA_{\mu}\exp{\left\{-\frac14\int d^4x F_{\mu\nu}^{a2}-\int d^4x
\psi^\dagger(-i\hat \partial -im -\hat A)\psi \right\}}
\label{SqQ}
\ee
$$
\times\psi^\dagger(x) S_{\bar Q} (x,y|A)\psi(y),
$$
where $S_{\bar Q} (x,y|A)$ is the propagator of the static antiquark placed at the origin.
For further analysis it is convenient to fix the modified Fock--Schwinger gauge \cite{FSg},
\be
\vec{x}\vec{A}(x_4,\vec{x})=0,\quad A_4(x_4,\vec{0})=0,
\label{5}
\ee
which ensures that the gluonic field vanishes at the trajectory of the static particle. As a result, the static antiquark decouples 
from the system, it is Green's function being simply
\be
S_{\bar Q}(x,y|A)=S_{\bar Q}(x,y)= i\frac{1-\gamma_4}{2} \theta (x_4-y_4) e^{-M(x_4-y_4)}+i\frac{1+\gamma_4}{2}\theta(y_4-x_4)e^{-M(y_4-x_4)}.
\label{SQ}
\ee

It is easy now to perform integration of the the gluonic field in Eq.~(\ref{SqQ}) to arrive at
\be
S_{q\bar Q}(x,y)=\frac{1}{N_C}\int D{\psi}D{\psi^\dagger}\exp{\left\{-\int d^4x L_{\rm eff}(\psi,\psi^\dagger)\right\}}\psi^\dagger(x) S_{\bar Q} (x,y)\psi(y),
\ee
with $L_{\rm eff}(\psi,\psi^\dagger)$ being the effective Lagrangian of the light quark moving in the field of the static antiquark source:
$$
\int d^4x L_{\rm eff}(\psi,\psi^\dagger)=\int d^4x\psi^\dagger_{\alpha}(x)(-i\hat\partial -im)\psi^{\alpha}(x)+ 
\int d^4x\psi^\dagger_{\alpha}(x)\gamma_{\mu}\psi^{\beta}(x)\langle {A_{\mu}}^{\alpha}_{\beta}\rangle
$$
\be
+\frac{1}{2}\int d^4x_1d^4x_2\psi^\dagger_{\alpha_1}(x_1)\gamma_{\mu_1}\psi^{\beta_1}(x_1)
\psi^\dagger_{\alpha_2}(x_2)\gamma_{\mu_2}\psi^{\beta_2}(x_2)\langle
{A_{\mu_1}}_{\beta_1}^{\alpha_1}(x_1){A_{\mu_2}}_{\beta_2}^{\alpha_2}(x_2)\rangle+
\ldots,
\label{8}
\ee
where all $\alpha$'s and $\beta$'s are fundamental colour indices, and the irreducible correlators 
$\langle {A_{\mu_1}}_{\beta_1}^{\alpha_1}(x_1)\ldots {A_{\mu_n}}_{\beta_n}^{\alpha_n}(x_n)\rangle$ of all orders enter. The 
first correlator, $\langle {A_{\mu}}^{\alpha}_{\beta}\rangle$, obviously vanishes due to the gauge and
Lorentz invariances of the vacuum. In what follows we assume the Gaussian dominance to take place in the QCD vacuum and thus we keep only the 
bilocal correlator 
\be
\langle{A_{\mu}}_{\beta}^{\alpha}(x){A_{\nu}}_{\delta}^{\gamma}(y)\rangle\equiv
2(\lambda_a)_{\beta}^{\alpha}(\lambda_a)_{\delta}^{\gamma}K_{\mu\nu}(x,y),
\ee
and neglect contributions of all higher correlators\ftnote{1}{This approximation leads to an exact Casimir scaling, 
that is the ratio of any two potentials between static sources in different representations of the colour group is given by the ratio of the
Casimir operators evaluated for the given two representations \cite{cs1}. The Casimir scaling was tested on the lattice \cite{casscal} and it was found to
manifest itself with a very high accuracy, which evidences a suppression of higher gluonic correlators as compared to the Gaussian one and
thus justifies the approximation made above.}. Then, using the relation $(\lambda_a)_{\beta}^{\alpha}(\lambda_a)_{\delta}^{\gamma}=\frac12
\delta_{\delta}^{\alpha}\delta_{\beta}^{\gamma}-\frac{1}{2N_C}\delta_{\beta}^{\alpha}\delta_{\delta}^{\gamma}$ and considering the
large--$N_C$ limit, we rewrite the effective light--quark Lagrangian in the form:
\be
L_{\rm eff}(\psi,\psi^\dagger)=\psi^\dagger_{\alpha}(x)(-i\hat\partial -im)\psi^{\alpha}(x)
+\frac{1}{2}\int d^4y\; \psi^\dagger_{\alpha}(x)\gamma_{\mu}\psi^{\beta}(x)\psi^\dagger_{\beta}(y)\gamma_{\nu}\psi^{\alpha}(y)
K_{\mu\nu}(x,y),
\ee
which leads to the Schwinger--Dyson-type equation \cite{hlya},
\begin{eqnarray}
\ds (-i\hat{\partial}_x-im)S(x,y)&-&i\int d^4zM(x,z)S(z,y)=\delta^{(4)}(x-y),\nonumber\\[-3mm]
\label{DS4}\\[-3mm]
\ds -iM(x,z)&=&K_{\mu\nu}(x,z)\gamma_{\mu}S(x,z)\gamma_{\nu},\nonumber
\end{eqnarray}
for the colour trace of the light--quark Green's function $S(x,y)=\frac{1}{N_C}\langle\psi^{\beta}(x)\psi^\dagger_{\beta}(y)\rangle$.

In order to evaluate the quark kernel $K_{\mu\nu}(x,y)$ we notice that a celebrated property of the radial gauges (gauge (\ref{5})
obviously belonging to this class) is a possibility to express the gluonic field $A$ in terms of the field strength tensor $F$. For the gauge
(\ref{5}) such a relation reads: 
\begin{eqnarray}
\ds A^a_4(x_4,\vec x)&=&\int_0^1 d\alpha x_i F^a_{i4}(x_4,\alpha\vec{x})\nonumber\\[-3mm]
\label{AF}\\[-3mm]
\ds A^a_i(x_4,\vec x)&=&\int_0^1\alpha x_k F^a_{ki}(x_4,\alpha\vec{x}) d\alpha,\quad i=1,2,3,\nonumber
\end{eqnarray}
so that the kernel $K_{\mu\nu}$ can be expressed in terms of field strength 
correlator $\langle F^a_{\mu\nu}(x)F^b_{\lambda\rho}(y)\rangle$, for which we use the
parametrisation \cite{VCM}:
\be
\langle F^a_{\mu\nu}(x)F^b_{\lambda\rho}(y)\rangle=\frac{\delta^{ab}}{N_C^2-1}D(x-y)
(\delta_{\mu\lambda}\delta_{\nu\rho}-\delta_{\mu\rho}\delta_{\nu\lambda})+
\Delta^{(1)},
\label{15}
\ee
where the second term $\Delta^{(1)}$ is a full derivative and it does not contribute to confinement and therefore will not be considered
below. 
The profile function $D(x-y)$ decreases in all directions of the Euclidean space, and this decrease is governed by the gluonic correlation
length $T_g$. Lattice simulations give rather small values of $T_g\approx 0.2\div 0.3$ fm \cite{lattice,km0}, so that the profile $D(x-y)$ 
has the support at close point $y\approx x$. The term proportional to $D(x-y)$ in (\ref{15}) contributes to 
the area law with the string tension \cite{VCM}
\be
\sigma=2\int_0^\infty d\tau\int_0^\infty d\lambda D(\tau,\lambda).
\label{sigma}
\ee

Then, in view of Eqs.~(\ref{AF}) and (\ref{15}), the quark kernel $K_{\mu\nu}(x,y)=K_{\mu\nu}(x_4-y_4,\vx,\vy)$ is found to be 
($\tau=x_4-y_4$)
\be
\left\{
\begin{array}{l}
K_{44}(\tau,\vx,\vy)=\ds(\vx\vy)\int_0^1d\alpha\int_0^1 d\beta D(\tau,|\alpha\vx-\beta\vy|),\\[1mm]
K_{i4}(\tau,\vx,\vy)=\ds K_{4i}(\tau,\vx,\vy)=0,\\[1mm]
K_{ik}(\tau,\vx,\vy)=\ds((\vx\vy)\delta_{ik}-y_ix_k)\int_0^1\alpha d\alpha\int_0^1 \beta d\beta
D(\tau,|\alpha\vx-\beta\vy|).
\end{array}
\right.
\label{kern1}
\ee

The Schwinger--Dyson-type equation (\ref{DS4}) is an essentially nonlinear equation. Its linearised form, with the Green's function $S(x,z)$
substituted by the free--quark Green's function $S_0(x,z)$ in the mass operator $M(x,z)$, can be used if the quark is heavy. 
This was done in Refs.~\cite{hl,hlus}, where the effective potential for the heavy quark interaction with the static antiquark source was
built, including the nonperturbative spin--orbit interaction. The leading correction to this potential due to the proper string dynamics 
was identified in Ref.~\cite{an0}. It was noticed in Ref.~\cite{hlus}, however, that the given linearisation procedure is selfconsistent only if the
product of the quark mass and the gluonic correlation length is large, $mT_g\gg 1$. In case $mT_g\ll 1$ the series of corrections to the
leading regime blows up and no conclusion concerning the dynamics of the system can be made \cite{hlus}.
This procedure is useless therefore for the purposes of the present paper which is aimed to consideration of the light (massless) quark 
with its effective mass generation due to the phenomenon of spontaneous breaking of chiral symmetry. Thus we study this case using a
different approach.

Below we use two simplifications:
i) we neglect the spatial part of the kernel $K_{ik}$
and ii) neglect corrections due to the finitness of the correlation length $T_g$.
The first approximation utilises the fact that, although
the spatial part of the kernel is important for the correct account of the QCD string rotation,
it is not decisive for the Lorentz nature of confinement, yielding only unnecessary complications.
The second approximation is justified in view of the results of the lattice simulations which give, as was mentioned before, quite 
small values of $T_g$.
The latter simplification allows us to approximate the kernel (\ref{kern1}) by an instantaneous kernel and thus,
for the Fourier transform of $K$ in time, to neglect its dependence on the energy,
\be
K_{44}(\omega,\vx,\vy)\equiv K(\omega,\vx,\vy)= K(\vx,\vy)=
(\vx\vy)\int_0^1d\alpha\int_0^1 d\beta \int_{-\infty}^{\infty} d\tau D(\tau,|\alpha\vx-\beta\vy|).
\label{kern2}
\ee
Then the mass operator can be written as
\be
M(x,y)=\delta(x_4-y_4)M(\vx,\vy),\quad M(\vx,\vy)=\frac12 K(\vx,\vy)\gamma_4\Lambda(\vx,\vy),
\label{Mop1}
\ee
where, following Ref.~\cite{hlya}, we introduced the quantity
\be
\Lambda(\vx,\vy)\equiv\sum_{n=-\infty}^\infty\psi_n(\vx){\rm sign}(n)\psi^\dagger_n(\vy)=
2i\int\frac{d\omega}{2\pi}S(\omega,\vx,\vy)\gamma_4=2iS(x_4-y_4,\vx,\vy){\gamma_4}_{|x_4=y_4},
\label{iS5}
\ee
which is convenient for studies of the Lorentz nature of the interquark interaction.
In addition, in the limit $T_g\to 0$, the profile function $D(\tau,\ld)$ takes a singular form
$D(\tau,\ld)=2\sigma\delta(\tau)\delta(\lambda)$ (see also Ref.~\cite{z} for the discussion of the singular limit of
some stochastic model). For such a profile one finds readily for the kernel (\ref{kern2}):
\be
K(\vx,\vy)=2\sigma(\vx\vy)\int_0^1d\alpha\int_0^1 d\beta\;\delta(|\alpha\vx-\beta\vy|).
\label{K4}
\ee
Evaluation of the integrals in Eq.~(\ref{K4}) is trivial and gives
\be
K(\vx,\vy)=2\sigma {\rm min}(|\vx|,|\vy|)=\sigma (|\vx|+|\vy|-|\vx-\vy|),
\label{K4v4}
\ee
if the vectors $\vx$ and $\vy$ are collinear (the kernel vanishes otherwise, as required by the delta--function in Eq.~(\ref{K4})). 
The requirement of collinearity of the vectors $\vx$ and $\vy$ ensures that the interaction between the light
quark and the static antiquark is due to an infinitely thin string --- a two--dimensional object embedded
into the four--dimensional space. In order to proceed we simplify the form of the kernel (\ref{K4v4}) 
and relax the constraint of collinearity. Thus we approximate the kernel as
\be
K(\vx,\vy)=\sigma(|\vx|+|\vy|-|\vx-\vy|).
\label{K4v3}
\ee

The quark kernel (\ref{K4v3}) possesses a number of important properties:
\begin{itemize}
\item it allows us to pass over back, to Minkowski space --- it will be used from now onward in this paper;
\item it admits a clear interpretation. Indeed, the kernel can be split into two parts: 
the local part $-\sigma|\vx-\vy|$ which is
responsible for the selfinteraction of the light quark, and the nonlocal part $\sigma(|\vx|+|\vy|)$ which 
describes the interaction of the
light quark with the static source. Such a form of the kernel is a consequence of the gauge condition (\ref{5}) 
which decouples the static particle from the system and brings all the information about the antiquark to
the kernel $K(\vx,\vy)$;
\item it admits a natural generalisation from the linearly rising potential to the potential of a generic 
form $V(r)$. In order to emphasise this important property we keep the potential as $V(r)$ in all formulae
below. Nevertheless, every time we need to specify the form of the potential, the linear confinement is understood, as the
most phenomenologically justified candidate;
\item with the kernel (\ref{K4v3}) we establish a link between the VCM and the GNJL models for QCD 
with instantaneous quark kernels 
which have a long history in the literature \cite{pqm,BR0} and which can be viewed as nonlocal divergence--free
generalisations of the Nambu--Jone-Lasinio model \cite{NJL} (recent exhaustive studies of the mesonic 
spectrum in this model can be found in Ref.~\cite{GW0}). Below we employ the chiral angle approach which
is widely used in such potential quark models.
\end{itemize}

Combining Eqs.~(\ref{DS4}), (\ref{Mop1}), (\ref{iS5}), and (\ref{K4v3}) together 
we arrive at the Schwinger--Dyson-type equation for the heavy--light quarkonium in the form:
\be
\left(-i\gamma_0\frac{\partial}{\partial t}+i\vg\frac{\partial}{\partial \vx}-m\right)
S(t,\vx,\vy)-\int d^3z M(\vx,\vz)S(t,\vz,\vy)=\delta(t)\delta^{(3)}(\vx-\vy),
\label{DS5}
\ee
where 
\be
M(\vx,\vz)=-\frac{i}{2}K(\vx,\vz)\gamma_0\Lambda(\vx,\vz),\quad\Lambda(\vx,\vz)=2i\int\frac{d\omega}{2\pi}S(\omega,\vx,\vy)\gamma_0,
\label{Mop01}
\ee
and the quark kernel is
\be
K(\vx,\vy)=V(|\vx|)+V(|\vy|)-V(|\vx-\vy|).
\ee
In the next section we study the properties of this equation.

\section{The Lorentz nature of confinement}

Let us investigate the properties of the quantity $\Lambda(\vx,\vy)$ and demonstrate the way it
acquires the contribution with the matrix structure $\propto\gamma_0$, since exactly this phenomenon 
constitutes spontaneous breaking of chiral symmetry. 

It was argued in Ref.~\cite{hlya} that Eq.~(\ref{DS4}) admits linearisation via the substitution 
\be
\Lambda(\vx,\vy)\approx\gamma_0\delta^{(3)}(\vx-\vy)+\ldots,
\label{Lm}
\ee
where the ellipsis denotes subleading at large distances terms. 
Such a substitution was justified then by an explicit computation of this
quantity using the spectrum of the resulting linearised Eq.~(\ref{DS5}). 
In Appendix~\ref{A} we give some details of the derivation of Eq.~(\ref{Lm}), taken from Ref.~\cite{hlya}.
The scalar Lorentz nature of the effective interaction follows immediately from the form of 
Eq.~(\ref{DS5}) and the matrix structure of $\Lambda(\vx,\vy)$ given in Eq.~(\ref{Lm}). 
We conclude therefore that Eq.~(\ref{DS5}) does admit a solution given by the 
scalar interaction generated in a selfconsistent manner.
Let us have an insight into the details of this selfconsistent generation of the scalar effective 
interquark interaction.

The separation of the kernel into the local and nonlocal parts mentioned above hints us the way to proceed. 
Indeed, let us consider the local part of the kernel first and omit the nonlocal part. Then Eq.~(\ref{DS4}) reduces to the Dyson equation for
the light--quark propagator
\be
(\gamma_0 p_0-\vg\vpp-m-\Sigma(\vpp))S(p_0,\vpp)=1,
\label{eqS0}
\ee
where the mass operator $\Sigma(\vpp)$ does not depend on the energy due to the instantaneous nature of the interaction and can be evaluated
as
\be
\Sigma(\vec{p})=-i\int\frac{d^4k}{(2\pi)^4}V(\vec{p}-\vec{k})\gamma_0 S(k_0,\vec{k})\gamma_0.
\label{Sigma01}
\ee 

\begin{figure}[t]
\begin{center}
\epsfig{file=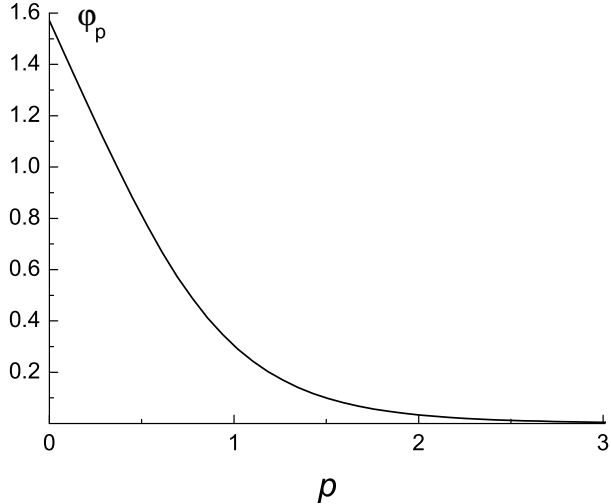,width=8cm} 
\caption{The profile of the solution to the mass--gap Eq.~(\ref{cha}) with $m=0$ and $V(r)=\sigma r$. The
momentum $p$ is given in the units of $\sqrt{\sigma}$.}\label{figvp}
\end{center}
\end{figure}

Equations (\ref{eqS0}) and (\ref{Sigma01}) together lead to the selfconsistent nonlinear equation for the quark mass operator \cite{pqm},
\be
\Sigma(\vec{p})=-i\int\frac{d^4k}{(2\pi)^4}V(\vec{p}-\vec{k})\gamma_0\frac{1}{\gamma_0
k_0-\vg\vk-m-\Sigma(\vk)}\gamma_0.
\label{Sigma03} 
\ee 
If we parametrise now the mass operator in the form:
\be
\Sigma(\vec{p})=[A_p-m]+(\vec{\gamma}\hat{\vec{p}})[B_p-p],
\label{SiAB} 
\ee 
with $A_p$ and $B_p$ being two auxiliary functions, then Eq.~(\ref{Sigma03}) gives the selfconsistency 
conditions for such a parametrisation,
\be
A_p=m+\frac{1}{2}\int\frac{d^3k}{(2\pi)^3}V(\vec{p}-\vec{k})\sin\vp_k,\quad
B_p=p+\frac{1}{2}\int \frac{d^3k}{(2\pi)^3}\;(\hat{\vec{p}}\hat{\vec{k}})V(\vec{p}-\vec{k})\cos\vp_k,
\label{AB} 
\ee 
where the angle $\vp_p$ --- known as the chiral angle --- is introduced to obey the condition
\be
A_p\cos\vp_p=B_p\sin\vp_p,
\label{cha} 
\ee
which, given the relations (\ref{AB}), plays the role of the mass--gap equation for the chiral angle.
Historically the chiral angle $\vp_p$ is defined such that $\vp_p(p=0)=\frac{\pi}{2}$ and 
$\vp_p(p\to\infty)=0$. In Fig.~\ref{figvp} we plot the chiral angle --- solution to the mass--gap
Eq.~(\ref{cha}) for the linear confinement. The interested reader can find the details of the chiral angle
formalism in Refs.~\cite{pqm,BR0,rep}. A comprehensive analysis of the properties of the mass--gap equation and
its solutions for various powerlike potentials is given in Ref.~\cite{BN0}.

It is an easy task now to evaluate the quantity $\Lambda(\vpp,\vq)$, which is the double Fourier transform of
$\Lambda(\vx,\vy)$:
\be
\Lambda(\vpp,\vq)=2i\int\frac{d\omega}{2\pi}S(\omega,\vpp,\vq)\gamma_0=(2\pi)^3\delta^{(3)}(\vpp-\vq)U_p,
\label{iS4}
\ee
where
\be
U_p=\beta\sin\vp_p-(\vec{\alpha}\hat{\vpp})\cos\vp_p,\quad \beta=\gamma_0,\quad\vec{\alpha}=\gamma_0\vg.
\label{Up4}
\ee

Equation (\ref{Up4}) gives the answer to the question on the Lorentz nature of confinement in the
heavy--light quarkonium. Indeed, for low--lying states with the relative momentum $p$ being small, the chiral
angle $\vp_p$ is close to $\pi/2$, so that the matrix $U_p=\beta$ and this immediately leads one to
Eq.~(\ref{Lm}). Notice that the contribution to $\Lambda(\vx,\vy)$ proportional to the matrix $\gamma_0$
appeared entirely due to chiral symmetry breaking described in terms of the nontrivial chiral angle $\vp_p$ --- see Fig.~\ref{figvp}
($\vp_p\equiv 0$ for the massless quark and without chiral symmetry breaking\ftnote{2}{The situation is trivial for the heavy quark when
chiral symmetry is broken explicitly. Indeed, in this case the
chiral angle acquires the contribution $\arctan(m/p)$ and thus $\vp_p\approx\pi/2$ for $p\ll m$.}). This is the regime
found in Ref.~\cite{hlya} and mentioned in the beginning of this section. In the next section we demonstrate
that one needs exactly this regime to realise in order to be able to describe the quarkonium using the
Salpeter equation. The opposite situation of the vanishing chiral angle, which realises for highly excited bound states, is discussed in
detail in Ref.~\cite{knr}.

Obviously, inclusion of the spatial part of the kernel (\ref{kern1}) as well as relaxing other simplifying
assumptions made in course of this section do not change the main conclusion of this section  ---
namely that, as soon as chiral symmetry is broken spontaneously, an effective scalar interaction appears in a selfconsistent manner.

\section{Scalar confinement and the Salpeter bound--state equation}

In the previous section we considered the approach to the heavy--light quark--antiquark system based on the Schwinger--Dyson-type 
Eq.~(\ref{DS4}). An alternative approach to (heavy--light) quarkonia, based on the spinless Salpeter equation,
\be
[\sqrt{p^2+m^2}+\sigma r]\psi=E\psi,
\label{Salp}
\ee
is also celebrated in the literature (see, for example, Ref.~\cite{Olsson} 
and Appendix~\ref{B} for the derivation of Eq.~(\ref{Salp}) in the formalism of the QCD string with quarks at the
ends \cite{DKS}, which is also based on the VCM \cite{VCM}). The purpose of the present section is to demonstrate that Eq.~(\ref{Salp}) and its more
sophisticated versions, like the Hamiltonian of the QCD string with quarks at the ends (see Appendix~\ref{B}), 
are consistent with the bound--state Eq.~(\ref{FW4}) under
the assumption of the scalar confinement dominance in the effective interquark interaction.

We return now to the full Eq.~(\ref{DS5}) and rewrite it in the form of the bound--state
equation for the bispinor wave function $\Psi$,
\be
(\vec{\alpha}\hat{\vpp}+\beta m)\Psi(\vx)+\beta\int d^3z M(\vx,\vz)\Psi(\vz)=E\Psi(\vx),
\label{DS6}
\ee
with the nonlocal part of the kernel included together with the local one. 
Notice that the heavy--light bound--state equation in the form Eq.~(\ref{DS6}) can be derived independently 
in the formalism of the GNJL quark models \cite{knr}. 

Passing over to the momentum space and using the mass--gap equation in the form:
\be 
E_pU_p={\vec \alpha}\vpp+\beta
m+\frac12\int\frac{d^3k}{(2\pi)^3} V(\vpp-\vk)U_k, 
\ee
one can rewrite Eq.~(\ref{DS6}) as 
\be
E_pU_p\Psi(\vpp)+\frac12\int\frac{d^3k}{(2\pi)^3}V(\vpp-\vk)[U_p+U_k]\Psi(\vk)=E\Psi(\vpp),
\label{Se10} 
\ee 
where the quantity $E_p=A_p\sin\vp_p+B_p\cos\vp_p$ is the full quark dispersive law which substitutes the free dispersion
$\sqrt{\vec{p}^2+m^2}$ and which appears as a result of the quark selfinteraction.

Equation~(\ref{Se10}) is subject to a Foldy--Wouthuysen transformation, which was built in a closed form in Ref.~\cite{knr}. The
corresponding operator is:
\be
T_p=\exp{\left[-\frac12(\vec{\gamma}\hat{\vec{p}})\left(\frac{\pi}{2}-\vp_p\right)\right]},\quad 
\Psi(\vec{p})=T_p{\psi(\vec{p})\choose 0},
\label{Tpop}
\ee
and the resulting Shr{\" o}dingerlike equation, which stems from Eq.~(\ref{Se10}) after the Foldy--Wouthuysen transformation with the
operator (\ref{Tpop}), reads:
\be
E_p\psi(\vec{p})+\int\frac{d^3k}{(2\pi)^3}V(\vpp-\vk)\left[C_pC_k+
({\vec \sigma}\hat{\vpp})({\vec
\sigma}\hat{\vk})S_pS_k\right]\psi(\vec{k})=E\psi(\vec{p}),
\label{FW4}
\ee
where we used the shorthand notations $C_p=\cos\frac12(\frac{\pi}{2}-\vp_p)$ and $S_p=\sin\frac12(\frac{\pi}{2}-\vp_p)$;
$\vec{\sigma}$ are Pauli matrices, and $\hat{\vpp}$ and $\hat{\vk}$ are unity vectors for $\vpp$ and $\vk$, respectively. 

Let us consider Eq.~(\ref{FW4}) in the regime $\vp_p\approx\frac{\pi}{2}$. Then $C_p=1$, $S_p=0$ and the interaction term in Eq.~(\ref{FW4})
reduces to the plain potential $\sigma r$, in coordinate space. Then the resulting equation reads:
\be
[E_p+\sigma r]\psi=E\psi,
\label{Salpm}
\ee
where $E_p$ plays the role of the kinetic energy operator for the quark. This equation has the form of the Salpeter Eq.~(\ref{Salp}).
Notice, however, that it is not always
sufficient to keep the kinetic term for light quarks in the form of the free--particle energy $\sqrt{\vec{p}^2+m^2}$, as in Eq.~(\ref{Salp}).
Strictly speaking, the dispersive law of the light quark $E_p$ is generated dynamically and has to be treated with care. This is especially
important for small interquark momenta, where $E_p$ can even become negative.
As a result, the lowest states in the spectrum --- the
pions and the kaons --- cannot be described using the simple Eq.~(\ref{Salp}) and the like. 
These states are to be considered using either the full Bethe--Salpeter
equation with the two--component mesonic wave function \cite{pqm,rep} or in the framework of the full
Schwinger--Dyson-type equation, similar to the heavy--light Eq.~(\ref{DS4}) \cite{yapion,19}.
A progress in adapting the Salpeter equation based approach to description of lightest mesons was achieved in Ref.~\cite{matrix} 
in the framework of a matrix Hamiltonian technique.
Apart from the aforementioned problem with the pions and kaons, the quark dispersive law $E_p$ in Eq.~(\ref{Salpm}) can be substituted, with
a good accuracy, by the free--quark energy, so that the Salpeter equation (\ref{Salp}) is readily reproduced.

In addition, as seen from Eq.~(\ref{FW4}), the simple Salpeter Eq.~(\ref{Salp}) has to meet certain problems for highly excited states as
well. Indeed, the relative interquark momentum is large in excited mesons, so that the chiral angle vanishes asymptotically (see
Fig.~\ref{figvp}). As a result, the
effective interaction in Eq.~(\ref{FW4}) becomes vectorial (see Eq.~(\ref{FW4}) with $C_p=S_p=1/\sqrt{2}$) and it does not reduce to a plain
potential anymore \cite{knr}.

We conclude this chapter stating that, contrary to naive expectations (potential is added to the energy), 
the form of the Salpeter Eq.~(\ref{Salp}) does not suggest that the
effective interquark interaction in the meson is vectorial. Furthermore, we demonstrate that this equation arises naturally from the full
Schwinger--Dyson-type equation for the heavy--light quarkonium under the assumption that chiral symmetry is broken (explicitly or spontaneously)
and, as a result, the effective scalar interquark interaction is generated in a selfconsistent manner. 

\section{Conclusions}

In this paper we address the problem of the Lorentz nature of confinement in QCD. We 
consider a heavy--light quark--antiquark system as a testground and
exploit the Schwinger--Dyson-type equation derived for the Green's function of such a system using the VCM. 
We demonstrate explicitly that the 
stringlike picture of the interquark interaction at large distances (in the form of the Salpeter equation (\ref{Salp})) 
appears due to chiral symmetry breaking. 
In particular, we prove that the Salpeter equation (\ref{Salp}) appears selfconsistently in the
Schwinger--Dyson approach to the heavy--light quarkonium if chiral symmetry is broken, explicitly or spontaneously, and the effectively
generated scalar potential
dominates in the effective interquark interaction.
This implies that the genuine Lorentz nature of the confining interaction in this Salpeter equation (as well as in the Hamiltonian of the 
QCD string with quarks at the ends) is scalar.  
This is the main result of this work. This solves the problem of the Klein paradox which
is known to operate for systems with vectorial interaction growing with the distance. 
We conclude that there is no room for such a problem in QCD. 

The reported result is robust since it is only based on quite a general consideration and is stable across the whole variety of 
quark kernels. Furthermore, our conclusions acquire additional support from the
fact that exactly the same bound--state equation for the heavy--light quarkonium can be derived independently in the framework of the GNJL
quark models which have a long history in the literature and are known to give deep insight into physics of chiral symmetry breaking.

Finally, let us mention that for light quarks and without spontaneous breaking of chiral symmetry one would have a vanishing chiral angle
and, consequently, no effective scalar interaction. This situation is believed to realise in QCD above the temperature of the
chiral symmetry restoration transition or for highly excited hadrons (see Ref.~\cite{G3} for a review). 
Properties of the interquark interaction in these situations deserves a special investigation and will be subject
for future publications (see, for example, Ref.~\cite{nsnew}).

\begin{acknowledgments}

This work was supported by the Federal Agency for Atomic Energy of Russian Federation and by the Presidential grant for leading scientific
schools NSh-843.2006.2. Yu.S. acknowledges support of the RFFI grant 06--02--17012.
A.N. is also supported via grants DFG-436 RUS 113/820/0-1(R), RFFI-05-02-04012-NNIOa, and PTDC/FIS/70843/2006-Fi\-si\-ca.
\end{acknowledgments}

\appendix

\section{Derivation of Eq.~(\ref{Lm})}\label{A}

Following the approach suggested in Ref.~\cite{hlya}, we assume (and justify this assumption {\em a posteriori}) 
that the Schwinger--Dyson Eq.~(\ref{DS5}) possesses a solution which gives for the 
quark mass operator (see Eq.~(\ref{Mop01})) a form described by a local scalar ($U(\vx)$) and local vector ($V(\vx)$) potential,
\be
M(\vx,\vy)=\left[U(\vx)+\gamma_0 V(\vx)\right]\delta^{(3)}(\vx-\vy).
\label{Mxy}
\ee
Then one can rewrite Eq.~(\ref{DS5}) in the form of a Dirac equation for the wave function $\psi(\vx)$,
\be
(\vec{\alpha}\vpp+\beta [m+U(\vx)]+V(\vx))\psi(\vx)=E\psi(\vx)
\label{DDD}
\ee
or, in components ($\psi=\frac{1}{r}{\hphantom{i}G_n\Omega_{jlm}\choose iF_n\Omega_{jl'm}}$),
\be
\left\{
\begin{array}{l}
\ds\frac{dG_n}{dr}+\frac{\kappa}{r}G_n-(E_n+m+U-V)F_n=0\hphantom{.}\\[3mm]
\ds\frac{dF_n}{dr}-\frac{\kappa}{r}F_n+(E_n-m-U-V)G_n=0.
\end{array}
\right.
\label{FGD}
\ee

One can use a simple trick to guess the matrix structure of the function $\Lambda(\vx,\vy)$,
\be
\Lambda(\vx,\vy)\equiv\sum_{n=-\infty}^\infty\psi_n(\vx){\rm sign}(n)\psi^\dagger_n(\vy),
\ee
built with the help of the solutions to Eq.~(\ref{DDD}). Indeed, according to its definition,
$\Lambda(\vx,\vy)$ can be naturally split into two parts,
\be
\Lambda^{(V)}(\vx,\vy)=\Lambda^{(V)}_+(\vx,\vy)-\Lambda^{(V)}_-(\vx,\vy),
\label{LLL}
\ee
where $\pm$ stand for the summation over positive and negative eigenvalues, respectively. Also, for future convenience, we used the script
$(V)$. A similar decomposition is valid for the reversed sign of the vector interaction,
\be
\Lambda^{(-V)}(\vx,\vy)=\Lambda^{(-V)}_+(\vx,\vy)-\Lambda^{(-V)}_-(\vx,\vy).
\ee
Now we notice the following symmetry inherent to the system (\ref{FGD}): $(V,E_n,\kappa,G,F)\leftrightarrow (-V,-E_n,-\kappa,F,G)$ and find
\be
\Lambda^{(V)}_+\propto (G\;iF){G^*\choose -iF^*}=
\left(
\begin{array}{cc}
GG^*&\\
&FF^*
\end{array}
\right),\quad
\Lambda^{(-V)}_-\propto
\left(
\begin{array}{cc}
FF^*&\\
&GG^*
\end{array}
\right).
\ee
Hence one can rewrite Eq.~(\ref{LLL}) as
\be
\Lambda^{(V)}=\left[\Lambda^{(V)}_+-\Lambda^{(-V)}_-\right]+\left[\Lambda^{(-V)}_--\Lambda^{(V)}_-\right]=
\gamma_0\sum_{E_n>0}(GG^*-FF^*)+\delta\Lambda,
\label{dL}
\ee
where the correction $\delta\Lambda$ vanishes for $V=0$. Therefore, for a purely scalar confinement, the matrix structure of
$\Lambda(\vx,\vy)$ is, indeed, given by the matrix $\gamma_0$. In order to establish its spatial structure one can use the WKB calculation
performed in Ref.~\cite{hlya}. We omit the lengthy calculation which can be found in Ref.~\cite{hlya} and quote here the final result:
\be
\Lambda(\vx,\vy)\approx\gamma_0\frac{\sigma}{\pi^2\sqrt{xy}}K_0\left(\sigma\sqrt{xy}|x-y|\right)\delta(1-\cos\theta_{xy}),
\label{Lap}
\ee
where $K_0$ is the MacDonald function. It is easy to check that,
\be
\int d^3 y \Lambda(\vx,\vy)=\gamma_0,
\ee
and, therefore, for $|\vx|,|\vy|\gg\frac{1}{\sigma|\vx-\vy|}$, $\Lambda(\vx,\vy)$ can be approximated by the three--dimensional
delta--function peaked at $\vy=\vx$. Thus we arrive at Eq.~(\ref{Lm}). Moreover, for $V\neq 0$, the same WKB method reproduces
Eq.~(\ref{Lap}) and gives the decrease of the term
$\delta\Lambda$ in Eq.~(\ref{dL}) at large distances, so that Eq.~(\ref{Lm}) holds \cite{hlya}. 

\section{Rotating QCD string and the spinless Salpeter equation}\label{B}

In this appendix we give a brief derivation of Eq.~(\ref{Salp}) in the formalism of the QCD string with quarks at the ends which is also
derived in the framework of VCM. 
Following the method of Ref.~\cite{DKS}, we start from the in-- and out--states of the quark--antiquark meson,
\be
\Psi^{({\rm in, out})}_{q\bar q}(x,y|A)=\bar{\Psi}_{\bar q}(x)\Phi(x,y)\Psi_q(y),\quad
\Phi(x,y)=P\exp{\left(ig\int_{y}^{x}dz_{\mu}A_{\mu}^at^a\right)},
\ee
and build its Green's function,
\be
G_{q\bar q}=\langle\Psi_{q\bar q}^{({\rm out})}(\bar{x},\bar{y}|A)\Psi^{({\rm in})}_{q\bar q}(x,y|A)^\dagger\rangle_{q\bar{q}A}
=\langle {\rm Tr}S_q(\bar{x},x|A)\Phi(x,y)S_{\bar{q}}(y,\bar{y}|A)\Phi(\bar{y},\bar{x})\rangle_A,
\ee
where $S_{q}$ and $S_{\bar{q}}$ are the propagators of the quark and the antiquark, respectively, in the background gluonic field. 
Averaging over the background field can be done using the minimal area law assumption for the isolated Wilson loop,
\be
\left\langle {\rm Tr}P\exp{\left(ig\oint_Cdz_{\mu}A_{\mu}\right)}\right\rangle_A\sim \exp{(-\sigma S_{\rm min})},
\ee
which is usually assumed for the stochastic QCD vacuum (see, for example, Ref.~\cite{VCM}) and is
found on the lattice. Here $S_{\rm min}$ is the area
of the minimal surface swept by the quark and antiquark trajectories,
\be
S_{\rm min}=\int_0^Tdt\int_0^1d\beta\sqrt{(\dot{w}w')^2-\dot{w}^2w'^2},
\ee
where, for the profile function of the string $w_\mu(t,\beta)$, we adopt the straight--line ansatz:
\be
w_{\mu}(t,\beta)=\beta x_{1\mu}(t)+(1-\beta)x_{2\mu},
\ee
$x_{1,2}(t)$ being the four--coordinates of the quarks at the ends of the string. We choose to consider the system in the laboratory frame
and also to synchronise the quark times,
\be
x_{10}=x_{20}=t.
\ee
The resulting Lagrangian of the string reads:
\be
L_{\rm str}=-\sigma r\int_0^1d\beta\sqrt{1-[\vec{n}\times(\beta\dot{\vec{x}}_1+(1-\beta)\dot{\vec{x}}_2)]^2},
\label{L2}
\ee
where $\vec{r}=\vx_1-\vx_2$, $\vec{n}=\vec{r}/r$. This interaction Lagrangian is to be supplied by the quark kinetic terms 
$-m_1\sqrt{1-\dot{\vx}_1^2}-m_2\sqrt{1-\dot{\vx}_2^2}$.
Then, with the help of the auxiliary (einbein) field technique, used to get
rid of the square roots in the kinetic (the einbeins $\mu_{1,2}$ \cite{einbein}) and in the string term
(the continuous einbein $\nu(\beta)$ \cite{DKS}) in the Lagrangian (\ref{L2}), one can proceed to the Hamiltonian of the system (see
Ref.~\cite{DKS} for the details of the derivation),
\begin{eqnarray}
\ds H&=&\sum_{i=1}^2\left[\frac{p_r^2+m_i^2}{2\mu_i}+\frac{\mu_i}{2}\right]+\int^1_0d\beta\left[\frac{\sigma^2r^2}{2\nu}+
\frac{\nu}{2}\right]\nonumber\\[2mm]
\label{Hm}\\[-3mm]
\ds &+&\frac{\vec{L}^2}{2r^2[\mu_1{(1-\zeta)}^2+\mu_2{\zeta}^2+\int^1_0d\beta\nu{(\beta-\zeta)}^2]},\quad\zeta=\frac{\mu_1+\int^1_0d\beta\nu\beta}{\mu_1+\mu_2+\int^1_0d\beta\nu}.\nonumber
\end{eqnarray}
Extrema in the einbein fields are understood either in the Hamiltonian (\ref{Hm}) or, alternatively, in its spectrum. In the latter
case the einbein field method is a variety of the celebrated variational method in Quantum Mechanics.

Now, if the contribution of the string to the total inertia of the rotating system (denominator of the last, angular--momentum--dependent,
term in the Hamiltonian (\ref{Hm})) is neglected, then the extrema in all einbeins can be taken analytically yielding for the
Hamiltonian (this procedure is exact for $L=0$):
\be
H=\sqrt{\vpp^2+m_1^2}+\sqrt{\vpp^2+m_2^2}+\sigma r,
\label{H0}
\ee
or, in the one--particle limit ($m_1\equiv M\to\infty$, $m_2\equiv m$),
\be
H=\sqrt{\vpp^2+m^2}+\sigma r,
\label{Hm2}
\ee
where we omitted the infinite contribution of the static particle mass $M$. After a canonical quantisation of the Hamiltonian (\ref{Hm2})
we reproduce the spinless Salpeter Eq.~(\ref{Salp}). The role of the proper string dynamics in the Hamiltonian (\ref{Hm}) as well as
supplying the spinless Hamiltonian with spin--dependent terms are discussed in the literature --- see, for example, 
Ref.~\cite{MNS} and Refs.~\cite{int,Lisbon}, respectively. Discussion of the proper string dynamics in the formalism of the
Schwinger--Dyson-type Eq.~(\ref{DS4}) can be found in Ref.~\cite{an0}.
Calculations of various hadronic spectra in the framework of the Hamiltonian (\ref{Hm}) supplied by the perturbative exchange and 
by the spin--dependent terms 
demonstrate a good accuracy of the predictions (see, for example, recent results for the spectrum of
heavy--light $D$, $D_s$, $B$, and $B_s$ mesons \cite{DB}). Notice that, in the case of light quarks, 
the major contribution to the spectrum of the Hamiltonian (\ref{Hm}) comes from the confining QCD string. 
Therefore, this case can be referred to as the case of the \lq\lq heavy" string (as opposed to the case of heavy quarks when the proper 
string dynamics gives only small corrections).

\end{document}